\documentclass[twocolumn]{aastex631}

\usepackage{ae,aecompl}
\usepackage{graphicx}
\usepackage{amsmath}
\usepackage{amssymb}
\usepackage{hyperref}
\usepackage{multirow}
\usepackage{txfonts}
\usepackage{xcolor}

\shorttitle{Quasi-matter Bounce Cosmology in Light of Planck}
\shortauthors{Arab \& Khorasani }
\graphicspath{{./}{figures/}}

\begin{document}
\title{Quasi-matter Bounce Cosmology in Light of Planck}
\author[0000-0002-0786-7307]{Mohammad Arab}\thanks{Contact e-mail: \href{m.arab92@basu.ac.ir}{m.arab92@basu.ac.ir}}
\affiliation{ICRANet-Mazandaran, University of Mazandaran, P. O. Box 47416-95447, Babolsar, Iran}

\author[0000-0002-7470-7135]{Mohsen Khorasani}\thanks{Contact e-mail:\href{mKhorasani@shirazu.ac.ir}{mKhorasani@shirazu.ac.ir}}
\affiliation{Department of Physics, School of Science, Shiraz University, Shiraz 71454, Iran}
\affiliation{Biruni Observatory, School of Science, Shiraz University, Shiraz 71454, Iran}

\begin{abstract}
	We study quasi-matter bounce cosmology in light of Planck cosmic microwave background (CMB) angular anisotropy measurements along with  the   BICEP2/Keck Array data. We propose a new primordial scalar power  spectrum  by considering a linear approximation of the equation of state $w\cong w_0+\kappa(\eta-\eta_0)$ for  quasi-matter field in the contracting phase of the universe. Using this new primordial scalar power  spectrum, we constrain the zeroth-order approximation of the equation of state $w_0= -\,0.00340\pm 0.00044$ and first-order correction $10^{4} \zeta= -1.67^{+1.50}_{-0.83}$ at the $1\sigma$ confidence level by Planck temperature and polarization in combination with the BICEP2/Keck Array data in which $\zeta = 12\kappa/k_*$ with pivot scale $k_*$. The spectral index of scalar perturbations is determined to be $n_{\rm Bs}=0.9623\pm0.0055$ which lies 7$\sigma$ away from the scale-invariant primordial spectrum. We find scale dependency for $n_{\rm s}$ at the $1\sigma$ confidence level and a tighter constraint on  the  running of the spectral index  compared to $\Lambda$CDM+$\alpha_s$ cosmology. The running of the spectral index in quasi-matter bounce cosmology  is $\alpha_{\rm Bs}= \pi \zeta /2 c_{\rm s} = -\hspace{.5mm}0.0021 \pm  0.0016$ which is nonzero  at the $1.3\sigma$ level,  whereas in $\Lambda$CDM+$\alpha_{\rm s}$ it is nonzero at the $0.8\sigma$ level for Planck temperature and polarization data.  The sound speed of density fluctuations  of the quasi-matter field  at the crossing time is $c_{\rm s} = 0.097^{+0.037}_{-0.023}$, which is not a very small value in the contracting phase.
\end{abstract}
\section{introduction}
The 2018 release of the Planck cosmic microwave background (CMB) measurements determined that the primordial scalar perturbations are nearly scale invariant  \citep{Akrami:2018odb}. Its results improve the Planck 2015 consequences \citep{Ade:2015lrj} in which they are more consistent with a vanishing scale dependency of the spectral index in standard model of cosmology. These measurements support the main predictions of single-field inflationary models,  which supposed the universe to have undergone a brief period of extremely rapid expansion right after the big bang \citep{Guth:1980zm,Linde:1981mu}. 

Cosmic inflation is a reasonable approach developed during the early 1980s to solve several hot big bang cosmological scenario defects \citep{Brout:1977ix,Kazanas:1980tx,Starobinsky:1980te,Guth:1980zm,Albrecht:1982wi,Linde:1981mu,Linde:1983gd}.
Although the $\Lambda$CDM inflationary model continues to be the most straightforward viable paradigm of the very early universe,  its classical implementation from the point of view of general relativity leads to singularities that arisen from the well-known Hawking-Penrose singularity theorem \citep{Hawking:1969sw}. Furthermore, in most inflation models, cosmological fluctuations' origin and early evolution of the physical wavelengths of comoving scales occur in the trans-Planckian regime, where general relativity and quantum field theory break down \citep{Martin:2000xs,Jacobson:1999zk,Brandenberger:2016vhg}.

Singularity and trans-Planckian problems are taken into account as the main weaknesses of the inflationary cosmology \citep{Martin:2000xs}. However, these problems are avoided in several scenarios for the very early universe as alternatives to cosmological inflation. For instance, the pre-big-bang scenario \citep{Veneziano:1991ek,Gasperini:1992em},  ekpyrotic/cyclic scenario \citep{Khoury:2001wf,Lehners:2007ac}, Emergent universe \citep{Ellis:2002we,Brandenberger:1988aj,Cai:2012yf,Cai:2013rna} and bounce cosmology reviewed in \citep{Qiu:2013eoa,Battefeld:2014uga} are some other configurations of the very early universe. The dynamical behavior of these configurations can be described by one or more theories such as string theory \citep{Battefeld:2005av,Kounnas:2011fk}, quantum gravity \citep{AcaciodeBarros:1997gy,Thiemann:2006cf}, and/or modified gravity \citep{Elizalde:2019tee,Odintsov:2020zct}. 

In bounce cosmology, an initial contraction phase precedes the expansion of the universe, and a big bounce basically replaces the big bang singularity. Loop quantum cosmology (LQC), which will be assumed for the high-energy regime in this article,  is derived by quantizing Friedmann–Lemaître–Robertson–Walker (FLRW) spacetime using loop quantum gravity (LQG) ideas and techniques \citep{Smolin:2004sx,Rovelli:2011eq}. As the universe contracts coming to its Planck regime, when the space-time curvature gets close to the Planck scale, quantum gravity effects become considerable and lead to a connection between the contraction and expansion phases of the universe at the bounce point \citep{Ashtekar:2007tv,Bojowald:2008ik,Ashtekar:2011ni,Cailleteau:2012fy,Cai:2014zga,Amoros:2013nxa}. 
Notice that the trans-Planckian problem is also avoided because the wavelengths of the fluctuations we are interested in remain many orders of magnitude larger than the Planck length \citep{Brandenberger:2016vhg,Renevey:2020zdj}.

There are a large number of scenarios to produce a bounce in the very early universe. However, non-singular matter bounce scenarios have been specifically investigated because of their potential to provide a perfect fit to the recent and future observations \citep{Cai:2011tc,Lehners:2015mra,Cai:2014xxa,Cai:2016hea,Cai:2015vzv}. It has commonly been assumed that the scale-invariant spectrum of curvature fluctuations corresponding to the cosmological observations come from the quantum vacuum fluctuations that originally exit the Hubble radius in a matter-dominated epoch of the contracting phase \citep{Haro:2014wha,Cai:2012va}.

Different methods have been proposed to establish a matter-dominated period in these scenarios. One of them is the matter-Ekpyrotic bounce, in which a single scalar field with non-trivial potential and non-standard kinetic term leads to an Ekpyrotic contraction before a non-singular bounce \citep{Cai:2014bea}. Some authors have considered two scalar fields, one of which operates as an ordinary matter, and the other guarantees generating a non-singular bounce \citep{Cai:2013kja}. In this case, the matter-dominated contracting universe moves to an ekpyrotic contraction phase and then, from a non-singular bounce period, goes to the phase of fast-role expansion.
Another possible scenario is matter bounce inflation, where a matter-like contracting phase before the inflation generalizes the inflationary cosmology \citep{Lehners:2015mra,Xia:2014tda}. 

The last one, which we will focus on more than the other ones, is the quasi-matter bounce scenario proposed by
\citep{Elizalde:2014uba}. In this scenario, a quasi-matter phase realized by a slight deviation from the exact matter field replaces the matter phase of the contracting universe in order to solve the tilt problem. Similar to the definitions of slow-roll parameters in inflationary cosmology, they introduced a set of parameters for the scenario to describe the nearly matter-dominated phase.
They explain that there exists a duality between a nearly quasi-matter contraction phase and the quasi de Sitter regime in the inflationary expanding universe.

This scenario is also independently addressed in \citep{Cai:2014jla} in the context of $\Lambda$CDM model. According to assumptions provided in this model, a period of the contracting phase is dominated by a positive cosmological constant and cold dark matter.
Although this innovative model was a scalar field-free model, which has been the first study of this kind of scenario where cold dark matter with the cosmological constant is used instead of the scalar field, it produced too much positive running of the spectral index \citep{deHaro:2015wda}. In contrast, base on 2013 and 2015 Planck results \citep{Ade:2015lrj,Planck:2013jfk}, the running was provided as a slightly negative value.

Matter contraction with interacting dark energy was offered in \citep{Cai:2016hea} to improve the scenario by generating a slight red tilt with little positive running. The authors of \citep{Arab:2017gae} suggested using Hubble-rate-dependent dark energy as a quasi-matter in the contracting phase to obtain a slightly negative running of the spectral index. 

The present work aims to answer this question: how can we directly get preliminary information about the equation-of-state parameter before the bounce as the main parameter in a quasi-matter bounce scenario by recent observational data?
We investigate a new approach to obtain meaningful parameters by generalizing the analytical solution of perturbation equations.
We directly check  parameters related to the spectral index of scalar perturbations and its scale dependency of quasi-matter bounce scenarios using Planck measurement of CMB angular anisotropy in combination with the BICEP2/Keck Array. 

This paper is organized into three main sections: the first section provides a brief review of the analytical calculation of cosmological perturbation, the second one is dedicated to explanation of the quasi-matter bounce cosmology, and the parameter estimation is considered for section 3.

We use the reduced Planck mass unit system and also a flat FLRW metric with a positive signature.
\section{Analytical calculations of cosmological perturbation} \label{sec2}
Near the Planck scale, the universe obeys quantum gravity rules instead of the classical one. In the context of holonomy-corrected LQC, quantum dynamics of the very early universe is described by a set of effective equations \citep{Ashtekar:2006wn},
\begin{align}
H^2&=\frac {\rho}{3}(1-\frac{\rho}{\rho_{\rm c}}),
\label{3}\\ \dot H&=(\frac{1}{2} \rho-3H^2)(1+w), \label{5}
\end{align} 
where $\rho$ and $p$ are total density and pressure, respectively, and $w$ is effective equation of state  $w=P /\rho$ and the dot denotes the time derivative. In terms of conformal time $\eta$\ in which $d \eta = {dt}/{a} $,  effective equations are simply  
rewritten as
\begin{align}
\mathcal H^2 & = \frac {\rho}{3} a^2 \left( 1-\frac{\rho}{\rho_{\rm c}}\right) \label{conformalbackground1}, \\ 
\mathcal H^\prime & =\frac{\rho}{2}a^2(1+w)-\mathcal H^2(2+3w), \label{conformalbackground2}\\
\rho^\prime & = -3\mathcal H\rho  (1+w), \label{c1}
\end{align}
where $\mathcal H ={a^\prime}/a=a H$ is the conformal Hubble rate  and prime denotes derivative with respect to conformal time. 

Loop-quantum-corrected dynamics of scalar  perturbations on a spatially flat background space-time are introduced by one of the 
modified Mukhanov-Sasaki equations \citep{Cailleteau:2012fy} 
\begin{equation} \label{pertur}
v'' - {c_{\rm s}}^2 \, \left(1 - \frac{2 \rho}{\rho_{\rm c}} \right)
\nabla^2 v - \frac{z''}{z} v = 0,
\end{equation}
where 
\begin{equation}
z = \frac{a \, \sqrt{\rho+P}}{c_{\rm s} \, H}\label{zs},
\end{equation}
and $ v = z \, \mathcal R$, is scalar-gauge-invariant Mukhanov-Sasaki
variable, $c_{\rm s}$ is the sound speed of density fluctuations, and $\mathcal R$ is the comoving curvature perturbation.

In the first place, during the quasi-matter-dominated epoch, effective equations \ref{conformalbackground1}, \ref{conformalbackground2}, and \ref{c1}  reduce to the standard Friedmann equations
\begin{align}
H^2&=\frac{\rho}{3}, \\
\dot{H}&=-\frac{3}{2} H^2(1+w) \label{hdot}, \\
\dot{\rho}&=-3 H \rho (1+w)\label{fridmann}.
\end{align}
Therefore, in terms of conformal time, in contracting phase, far enough from the bounce,  Friedmann equations  can be written by
\begin{equation}\label{mHeq}
\mathcal H'=-\frac{1}{2} (3 w + 1) \mathcal H^2.
\end{equation} 

A quasi-matter epoch is a nearly matter-dominated phase (see the next section). 
For a matter-dominated ($w=0$) regime, primordial scalar perturbations are scale invariant (Harrison-Zel'dovich power spectrum) \citep{Wilson-Ewing:2012lmx}, and for a quasi-matter ($|w(\eta)| \ll 1$), they become scale dependent \citep{Elizalde:2014uba,deHaro:2015wda}. Obviously, after the $w$-constant assumption, which is considered due to slowly evolving the effective equation of state in (quasi-)matter field \citep{Cai:2014jla},  the most trivial approximation is 
$w_0 + \kappa(\eta -\eta_0)$. It leads directly to a different primordial power spectrum with a new function of the main variables $\kappa$, $w_0$ and $c_s$.
Thus, we use the linear approximation of the effective equation of state   $w \cong w_0 + \kappa(\eta-\eta_0) $ in the contracting phase of the universe to find a more comprehensive and exact solution of the perturbation equation in the quasi-matter period. Using this approximation, equation (\ref{mHeq}) gives   
\begin{equation}\label{mH} 
\mathcal H=\frac{4}{3 \kappa(\eta-\eta_0)^2+2(3w_0+1)(\eta-\eta_0)},
\end{equation}
where $\eta_0$ is an integration constant that is introduced to preserves
the Hubble parameter continued at the equality time $\eta_{\rm e}$, which denotes the final moment of quasi-matter contraction and the beginning of the radiation phase.
Because of the low energy and curvature in the quasi-matter-dominated epoch, we can use the  $z''/z \cong a''/a$ approximation in the perturbation equation \citep{Elizalde:2014uba}. Thus using equation (\ref{mH}), we will have 
\begin{equation}\label{appzz}
\frac{z''}{z}\cong \mathcal H^2 + \mathcal H \cong\frac{2-18 w_0}{(\eta-\eta_0)^2}-\frac{12 \kappa}{\eta-\eta_0}.
\end{equation}
In order to rewrite the perturbation equation in mathematical standard  form, it is necessary to  define  $\nu \cong 3/2 - 6 w_0$. Using equation (\ref{appzz}), Mukhanuv-Sasaki equation  in Fourier modes becomes 
\begin{equation} 
v''+{c_{\rm s}}^2 k^2 \left(1 -\frac{\nu^2 -\frac{1}{4}}{{c_{\rm s}}^2 k^2 (\eta-\eta_0)^2}+\frac{\xi(\eta-\eta_0)}{{c_{\rm s}}^2 k^2
	(\eta-\eta_0)^2} \right) v=0 , \label{prt}    
\end{equation}
where $\xi = 12 \kappa $. The solution is
\begin{align}
v_k\,= & J_1 \, W\left( \frac{-i \, \xi}{2 \, {c_{\rm s}} \, k},\,\nu\, , 2 \, i \, {c_{\rm s}} \, k \,  (\eta- \eta_0) \right)  + \nonumber 
\\ &J_2 \, M\left( \frac{-i \, \xi}{2 \, {c_{\rm s}} \, k},\,\nu\, , 2 \, i \, {c_{\rm s}} \, k \, (\eta- \eta_0) \right), \label{Whittaker}
\end{align}
in which $W$ and $M$  are Whittaker functions and $J_1$ and $J_2$ are constant.  The asymptotic behavior of our solution for  $|k(\eta-\eta_0)|\gg1$, requires it to be initially consistent with the quantum vacuum state before the bounce. \citep{Arab:2017gae},
$$J_1= \sqrt{\frac{1}{2 {c_{\rm s}} k}} \exp(\frac{\pi \xi }{4 {c_{\rm s}} k}), ~~~~~~~~~~~~ J_2=0. $$
For the modes which satisfy the long-wavelength limit condition $|k (\eta- \eta_0)| \ll 1$, equation (\ref{Whittaker}) gives
\begin{align}
v_k  \cong   \frac{-i}{2}\sqrt{\frac{1}{2 {c_{\rm s}} k}} \exp(\frac{\pi \xi }{4 {c_{\rm s}} k}) \, \, \left(\frac{1}{2}{c_{\rm s}} \, k \, ( \eta - \eta_0 ) \right)^{-1+6 w_0}.
\end{align}
We obtain this equation by noting that $w_0$ and $\xi/({c_{\rm s}} k)$ are very close to zero in crossing time.

In the second place, after the equality time $\eta = \eta_e $, the universe is dominated   by radiation, $w = 1/3$. { Note that there is no assumption of interaction between radiation and a quasi-matter field. Although the effective equation of state changes continuously between these different periods, assuming it behaves like a step function allows us to solve the equation of motion of $v_k$  analytically \citep{Cai:2014jla}.}  It is easy to see that continuity equation implies $\rho= \rho_0/a^4 $. Therefore, {in the limit of vanishing $\rho/\rho_c$},  Friedmann equations give
\begin{equation} a(\eta) = \sqrt{\rho_0 /3} \; \eta, 
~~~~~~~ \mathcal H= \frac{1}{\eta}. 
\end{equation}
They lead  the equation of motion  to a harmonic oscillator with the following exact solution
\begin{equation}
v_k = B_1\, sin\left( \frac{k \eta}{\sqrt{3}}\right)  + B_2 \,cos\left( \frac{k \eta}{\sqrt{3}}\right), \label{radiation perturbation}  
\end{equation}
where $B_1$ and $B_2$ are determined by imposing continuity in $v_k$ and $v'_k$  at $\eta_e$
\begin{align}
B_1 &= A_1  \sin\left( \frac{k \eta_e}{3}\right) + A_2 \, \frac{3}{k} \cos\left( \frac{k \eta_e}{3}\right), \nonumber \\
B_2 &= A_1 \cos\left( \frac{k \eta_e}{3}\right) - A_2\, \frac{3}{k}  \sin\left( \frac{k \eta_e}{3}\right) ,
\end{align}
in which $A_1$ and $ A_2$ are
\begin{align}
A_1 \approxeq & -2i \, \exp\left(\frac{\pi  \xi }{4 \text{c}_s k}\right)\, (2 \,{c_{\rm s}} k)^{6 w_0-\frac{3}{2}} \text{$(\eta_{\rm e}-\eta_0)$}^{6 w_0-1}, \nonumber \\
A_2 \approxeq &\,\, 2i \, \exp\left(\frac{\pi  \xi }{4 \text{c}_s k}\right)\, (2 \,{c_{\rm s}} k)^{6 w_0-\frac{3}{2}} \text{$(\eta_{\rm e}-\eta_0)$}^{6 w_0-2} .
\end{align}
It is necessary to note that while the universe is contracting to the bounce epoch, $\eta \rightarrow 0$, the first term in equation (\ref{radiation perturbation}) can be neglected  compared to the second one; thus, the $B_2$ term is dominated near the bounce. 

We suppose the universe is dominated by radiation at the bounce period. Hence, substituting $\rho = \rho_0/a^4$ in the LQC effective equation (\ref{5}), we obtain
\begin{equation}
a(t) = \,  \left(  \frac{4 {\rho_0} }{3} \, t^2 + \frac{{\rho_0}}{{\rho_{\rm c}}}\right)^{1/4}.
\end{equation}
Besides, The long-wavelength limit is still satisfied during the bounce. Thus, the second term of equation (\ref{pertur}) is dropped and the perturbation equation for $v_k$ is reduced to
\begin{equation}
\frac{v_k''}{v_k} - \frac{z''}{z} = 0.
\end{equation}
From equation (\ref{zs}) we can obtain the asymptotic behavior of the solution for $t \ll  -\sqrt{3/4 \, \,\rho_{\rm c}} $ ,
\begin{equation}
v_k \approxeq \,\left(2 \frac{\sqrt{2} {{\rho _0^{1/4}}}}{{3}^{1/4}} C_1+\frac{\sqrt{\pi }  {{{\rho_{\rm c}}^{1/4} }}\, \Gamma \left({1}/{4}\right)}{ 3\,^{1/4} \,6\sqrt{\text{2}{\rho_0 }}\, \Gamma \left({3}/{4}\right)} C_2\right)  \sqrt{t} -\frac{1 }{2\,\sqrt{{\rho_0}}} \,C_2 ,
\label{vbounce}	
\end{equation}
in which $C_1$ and $C_2$ are determined by the fact that this expression must be compatible with the pre-bounce radiation-dominated epoch where the quantum gravity effects do not still have a significant role. 
So, the prefactor of  the first term of equation (\ref{vbounce}) must be zero, and then
\begin{equation}
C_1= -\frac{\sqrt{{\pi }}  {{\rho_c}}^{1/4} \,\,\Gamma \left({1}/{4}\right)}{4 \sqrt{3} {\rho _0}^{3/4} \,\, \Gamma \left({3}/{4}\right)} \, C_2 \,,
\end{equation}
\begin{equation}
C_2 = - 2 B_2 \sqrt \rho_0 \,.
\end{equation}
Therefore, in the classical  post-bounce region where $t\gg\sqrt{3/4\rho_c}~,$ comoving curvature perturbation can be obtained as a time-independent quantity in zeroth-order approximation,
\begin{equation}\label{mR}
\mathcal R= \frac{v_k}{z}= \frac{1}{6} \sqrt{\frac{\pi}{3}}\left( \frac{\rho_c}{\rho_0}\right)^{{1}/{4}} \frac{\Gamma({1}/{4})}{\Gamma({3}/{4})}\; B_2 + \mathcal O(t^{-1/2}).
\end{equation}

\begin{figure*}
	\centering
	\includegraphics[width=8cm]{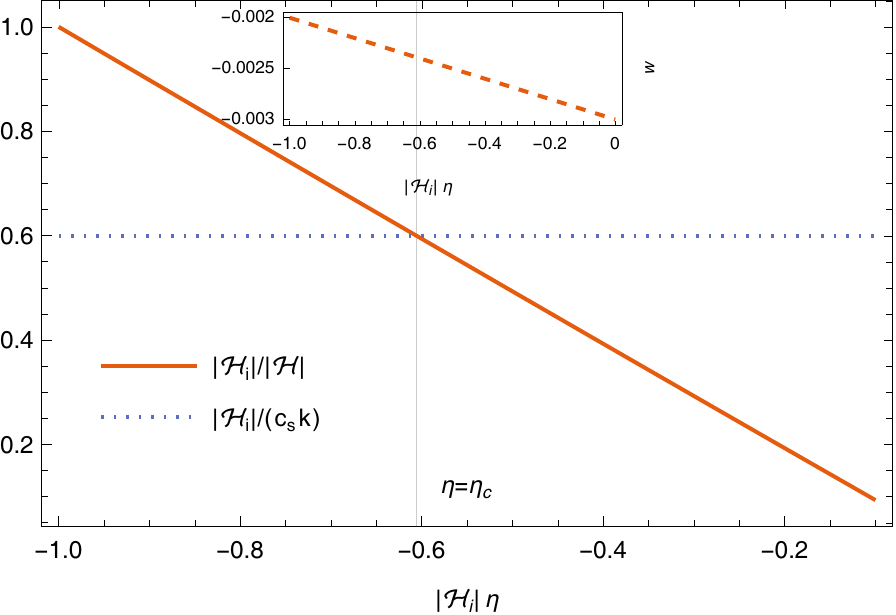}
	\caption{Evolution of the conformal Hubble radius (solid line) and the equation of state (dashed line) in the quasi-matter-dominated epoch in the contracting phase. $\eta_c$ is the crossing time at which the comoving sound wave length (dotted line) exits the horizon. $ w(\eta_c)$ is the amount of the effective equation of state $w$  at the crossing time.
		$\mathcal H_i$ is an initial conformal Hubble parameter. The slope of $w$ is exaggerated.}\label{CTplot}
\end{figure*}

Note that the nearly scale-invariant power spectrum is only achieved in the limit of $|k \, \eta_{\rm e}| \ll 1$ . this means that super-horizon perturbation modes in the quasi-matter contracting phase must be out of the sound Hubble radius at the equality time. It also results in an asymmetric horizon crossing before and after the bounce \citep{Cai:2014jla}. 
Thus, the power spectrum is deformed as
\begin{align} \label{power spectrum}
{\Delta_{\rm s}}^2 =& \frac{k^3 |\mathcal R|^2}{2 \pi^2}\\ \nonumber
=&  \frac{{({{c_{\rm s}} \eta_{\rm e}})^{-3+12 w_0}}}{108 \, \pi \, {\eta_{\rm e}}} \sqrt{\frac{\rho_{\rm c}}{\rho_0}  }\left( \frac{\Gamma(1/4)}{\Gamma(3/4)}\right) ^2  \exp\left(\frac{\pi \xi}{2 {c_{\rm s}} \, k
} \right) k^{12 w_0}.
\end{align}
Actually, the modified version of the nearly scale-invariant power spectrum is due to the linear approximation of the effective  equation of state at the time of the horizon crossing in the contracting phase of the universe; see Fig. \ref{CTplot}.
In a similar manner to scalar perturbations, long-wavelength modes of tensor perturbations (gravitational waves) can also come from the contracting phase through the bounce epoch. Nevertheless, compared with scalar perturbations, tensor perturbations are propagated with light speed. The effective equation of motion for tensor perturbation $h$ is given by
\begin{equation} \label{perturtensor}
\mu'' -  \, \left(1 - \frac{2 \rho}{\rho_{\rm c}} \right)
\nabla^2 \mu - \frac{z_T''}{z_T} \mu = 0,
\end{equation}
in which  $z_T = {a}/ \sqrt{1-2\rho/\rho_{\rm c}}$ and $\mu= {h} \, z_T  $ is the tensor gauge-invariant Mukhanov-Sasaki variable. The method used for tensor perturbation calculations is precisely similar to the scalar perturbation. We obtain
\begin{align} \label{Tensor_power spectrum}
{\Delta_{\rm h}}^2 =& \frac{k^3 |h|^2}{2 \pi^2}\\ \nonumber
=&  \frac{{2\,({\eta_{\rm e}})^{-3+12 w_0}}}{9 \, \pi \, {\eta_{\rm e}}} \sqrt{\frac{\rho_c}{\rho_0}  }\left( \frac{\Gamma(1/4)}{\Gamma(3/4)}\right) ^2  \exp\left(\frac{\pi \xi}{2 \, k
} \right) k^{12 w_0}.
\end{align}
In section \ref{PS} we will use this modified primordial scalar power spectrum to constrain the equation-of-state parameters. 

\section{quasi-matter bounce cosmology} \label{sec:power spectrum}
\noindent An extended matter bounce scenario can be explainable in light of a single scalar field in the quasi-matter-dominated contracting phase \citep{deHaro:2015wda}. A quasi-matter-dominated regime is actually a nearly matter-dominated phase in the contracting universe, where long-wavelength modes of perturbation exit the sound Hubble radius. This regime can be described by the cosmological constant plus cold dark matter  \citep{Cai:2014jla},  matter contraction with interacting dark energy \citep{Li:2016xjb}, Hubble-rate-dependent dark energy \citep{Arab:2017gae}, etc. 
A spatially homogeneous scalar  field $\varphi(t)$  with  potential $V(\varphi)$ is characterized by the energy  density $\rho_\varphi$ and  the pressure $ P_\varphi$
\begin{align}
&\rho_\varphi = \frac{1}{2} \dot{\varphi}^2+V(\varphi),\\ & P_\varphi = \frac{1}{2} \dot{\varphi}^2-V(\varphi),
\end{align} 
and by the time-dependent equation of state $w =P_\varphi/\rho_\varphi$.
In the FLRW  geometry we have 
\begin{align}
& \dot H = -\frac{1}{2} {\dot \varphi}^2,\\
&3 H^2= \frac{1}{2} \dot{\varphi}^2 +V(\varphi), \\
& \ddot{\varphi}+ 3 H  \dot{\varphi} + V,_{\varphi}=0,
\end{align} 
where $V,_{\varphi} \equiv{\partial V}/{\partial \varphi} $. For the case of the quasi-matter scalar-field-dominated phase $|w|\ll 1$, the background equations in conformal time become
\begin{align}\label{BG-Sfield-approx}
&3 \mathcal H^2 -{2} a^2 V \cong 0,\\ \label{BG-Sfield-approx2}
&3 \mathcal H \varphi' + 2 a^2 V,_\varphi\cong 0 \;.  
\end{align}
Thus, the equation-of-state parameter $ w$ turns to 
\begin{align}\label{wphi}
w = &\frac{1}{3} \left( \frac{V,_{\varphi}}{V}\right)^2 -1.
\end{align}
Utilizing the well-known slow-roll parameters $\epsilon$ and  $\delta$  helps us to compare bounce scenarios with inflationary models. The parameter\begin{equation}
\epsilon \equiv - \frac{\dot H}{H^2} = \frac{1}{2}\left(\frac{V_{,\varphi}}{V}\right)^2,
\end{equation}
performs a crucial role in the existence of a nonzero primordial tilt, in the inflationary cosmology $\epsilon \ll 1$, in the ekpyrotic/cyclic scenario $\epsilon \gg 1$, and for quasi-matter bounce $\epsilon \cong 3/2$.
For inflation, the spectral tilt is 
\begin{equation}
n_{\rm s} -1 = -4 \epsilon + 2 \delta,
\end{equation}  
where $\delta \equiv V_{,\varphi \varphi}/V$ \citep{Stewart:2001cd}. In the inflation these parameters satisfy slow-roll condition $\epsilon \ll 1$ and $\delta \ll 1$ which imply the  potential is flat. It is customary to define for the ekpyrotic/cyclic scenario  the quantity 
\begin{equation}
\bar \epsilon \equiv \frac{1}{2 \epsilon} = \left(\frac{V}{V_{,\varphi}}\right)^2, 
\end{equation}  
so the spectral tilt $
n_{\rm s}-1$ turns to   $-4 \bar \epsilon - 4 \bar \delta, $
in which $\bar \delta \equiv \left(V/V_{,\varphi}\right)_{,\varphi}$ \citep{Gratton:2003pe}.
In  this scenario these parameters satisfy ``fast-roll'' conditions $\bar\epsilon \ll 1$ and $\bar\delta \ll 1$ which translate into the requirement that the potential is steep \citep{Gratton:2003pe,Lehners:2007ac}.
In quasi-matter bounce cosmology, parameters  introduced by
\citep{Elizalde:2014uba} follow $\bar \epsilon \equiv w$  and $\bar \delta \equiv -\left(V_{,\varphi}/V\right)_{,\varphi}$, in which they  satisfy the conditions $\bar\epsilon \ll 1$ and $\bar\delta \ll 1$ in similar to the slow-roll parameters. 

Let us now turn to a brief review of the usual method to solve the perturbation equation for the quasi-matter bounce scenario with regard to the slow-roll parameters.
Using Eqs. (\ref{pertur}) and (\ref{zs}) with $c_{\rm s}z\equiv a(2\epsilon)^{1/2}$, we have
\begin{equation}\label{mkh z}
\frac{z^{\prime \prime}}{z} = {\mathcal H}^2 \left(2-\epsilon + \frac{3 \epsilon_{,N}}{2\epsilon}-\frac{\epsilon_{,N}}{2}\right),
\end{equation}
where $dN = d \ln a$ and the second-order derivative is dropped. Expression for $\mathcal H$ in terms of $\epsilon$ comes from definition ${\mathcal H}^{\prime} = {\mathcal H}^{2}(1-\epsilon)$, so successive integration by part gives 
\begin{equation}\label{mkh h inverse}
{\mathcal H}^{-1} = \eta(\epsilon -1)\left(1 -\frac{\epsilon_{,N}}{(\epsilon-1)^2}\right),
\end{equation}  
where terms of order $\epsilon_{,NN}$ and ${(\epsilon_{,N})}^2$ 
are ignored \citep{Lehners:2007ac,Lehners:2015mra}. Therefore, in the approximation that second-order derivatives are negligible, equation (\ref{mkh z}) turns to
\begin{equation}\label{mkh zs}
\frac{z^{\prime \prime}}{z} =  
\frac{2-\epsilon + \frac{3 \epsilon_{,N}}{2\epsilon}-\frac{\epsilon_{,N}}{2}+\frac{4\epsilon_{,N}}{(\epsilon-1)^2}-\frac{2\epsilon \epsilon_{,N}}{(\epsilon-1)^2}}
{(\epsilon -1)^2\eta ^2},
\end{equation} 
so in the limit $\rho/\rho_{\rm c} \rightarrow 0$, 
the solution of equation (\ref{pertur}) up to an irrelevant phase factor is the Hankel function $H^{(1)}_{\nu}(-c_{\rm s}k\eta)$ with index \citep{Lehners:2015mra} 
\begin{equation}\label{mkh nu index}
\nu = \frac{1}{2(\epsilon -1)} \left(3-\epsilon + \frac{\epsilon_{,N}}{\epsilon}+\frac{4(\epsilon-2)\epsilon_{,N}}{(\epsilon-3)(\epsilon-1)^2}\right).
\end{equation}
In quasi-matter bounce cosmology, the equation of state $|w| \ll 1$
implies $\epsilon \cong 3/2$. Therefore, it is appropriate to express scalar spectral index respect to $w$ instead of $\epsilon$. The result with neglecting the second-order derivative is 
\begin{align}\label{mkh spectral running}
n_{\rm s} - 1 & = 12 w + 9 \frac{{\rm d}w}{{\rm d}\ln k} \Big|_{{c_{\rm s}}k_*=|a_* H_*|}, \nonumber\\ 
\alpha_{\rm s} & \equiv \frac{{\rm d} n_{\rm s}}{{\rm d}\ln k} \Big|_{{c_{\rm s}}k_*=|a_* H_*|}= 12 \frac{{\rm d}w}{{\rm d}\ln k} \Big|_{{c_{\rm s}}k_*= |a_* H_*|}, 
\end{align}
in which ${\rm d}\ln k \Big|_{{c_{\rm s}}k_*= |a_* H_*|} = - (\epsilon-1){\rm d}N$  and $\alpha_{\rm s}$ is the running of the
scalar spectral index \citep{Lehners:2015mra}. $k_*$ denotes an arbitrary pivot scale, and $a_*$ and $H_*$  are the scale factor and Hubble parameter values at the crossing time (${c_{\rm s}} k_* = |a_* H_*|$).  The expressions in (\ref{mkh spectral running}) can be interpreted by the first-order expansion of the equation of state $12 w \cong 12 w_0+ \xi (\eta-\eta_0)$ in the quasi-matter contracting phase
\begin{align}
n_{\rm s} - 1 & = 12 w_{*} -  \frac{6 \xi}{(4+12w_*)\mathcal H_*} 
\label{mkh ns}\\\nonumber 
&\cong 12 w_{*}- \frac{3 \xi}{2 \mathcal H_*},
\\
\alpha_{\rm s} &= - \frac{8 \xi}{(4+12w_*)\mathcal H_*}\label{mkh alpha}\\ \nonumber
&\cong - \frac{2 \xi}{ \mathcal H_*} ,
\end{align}
where $w_*$ is the equation of state at the time $\eta_*$  at which the pivot scale $k_*$ crosses the sound Hubble radius, and $\mathcal H_*$ denotes the conformal  Hubble parameter at the crossing time. 
From equation (\ref{mkh alpha}), a nonzero slope $\kappa$ implies the existence of the running of the spectral index. 
The above relations and the nearly scale-invariant scalar power spectrum obtained in the previous section, equation (\ref{power spectrum}), motivate us to use directly quasi-matter characters ($w_0$, $\xi$) as the primary cosmological parameters in the procedure of observational constraints. In fact, there is no need to \textit{define} the extra parameter $\alpha_{\rm s}$ for investigating the scale dependency of scalar spectral index $n_{\rm s}$. As a final remark in this section, it is helpful to investigate one of the indirect results of our approach by giving an example.
Using Eqs.  \ref{BG-Sfield-approx}, \ref{BG-Sfield-approx2} and \ref{wphi}, we can easily find 
\begin{align}\label{w derivative}
\kappa = -2 \mathcal H (1+w)\left(\frac{V_{,\varphi}}{V} \right)_{,\varphi},
\end{align}
and from equation (\ref{wphi}) the potential is obtained as
\begin{align}\label{potential bounce}
V(\varphi)
\cong V_0 \exp\left(-\sqrt{3}\left[\varphi+f(\varphi)\right] \right) ,
\end{align}
in which  $f(\varphi)=\frac{1}{2}\int{{w(\varphi)}}{\rm d}\varphi$ is small compared to $\varphi$ and it shows that the universe contracts like a quasi-matter-dominated universe with a nearly zero pressure.  For example, if we suppose $w(\varphi) =  \beta \varphi^n$, we have 
\begin{align}\label{kappa}
\kappa = - n w \sqrt{3(1+w)}~\frac{\mathcal H}{\varphi}.
\end{align}
\begin{table*}
	\begin{center}
		\begin{tabular}{lc}
			\hline
			\hline
			Parameter   & Definition \\
			\hline
			$\overline{A}_s$  $\cdots$$\cdots$$\cdots$$\cdots$    & \hspace{1cm} Scalar power spectrum amplitude (at $k_{*} = 0.05 \hspace{1mm}\rm {Mpc^{-1}}$ )      \\
			$w_0$ $\cdots$$\cdots$$\cdots$$\cdots$ & \hspace{1cm}  Zero-order quasi-matter equation-of-state parameter (at $k_{*} = 0.05 \hspace{1mm}\rm {Mpc^{-1}}$ )      \\	
			$\xi$ $\cdots$$\cdots$$\cdots$$\cdots$$\cdot$ & \hspace{1cm}
			First-order quasi-matter equation-of-state parameter (at $k_{*} = 0.05 \hspace{1mm}\rm {Mpc^{-1}}$ )      \\
			${c_{\rm s}}$ $\cdots$$\cdots$$\cdots$$\cdots$ & \hspace{1cm} 
			Ratio of quasi matter field sound speed to light speed (at $k_{*} = 0.05 \hspace{1mm}\rm {Mpc^{-1}}$) 
			\\ 
			$\Omega_{\rm b} h^2$    $\cdots$$\cdots$$\cdots$  & \hspace{1cm}  Baryon density       \\
			$\Omega_{\rm c} h^2$   $\cdots$$\cdots$$\cdots$ & \hspace{1cm} Cold dark matter density      \\
			$\theta_{\rm{MC}}$  $\cdots$$\cdots$$\cdots$$\cdot$ & \hspace{1cm} 
			Angular size of sound horizon at last scattering      \\
			$\tau$ $\cdots$$\cdots$$\cdots$$\cdots$$\cdot$ & \hspace{1cm} Optical depth of reionized intergalactic medium  \\	
			\hline
		\end{tabular}
	\end{center}\caption{Primordial, baseline, and optional late-time cosmological parameters for quasi-matter bounce cosmology.} \label{mkh table 0}
\end{table*}
One can find some properties of the potential of the quasi-matter scalar field dominated in the contracting phase by using the linear approximation of the equation of state.  Equation (\ref{kappa}) at crossing time  turns to
\begin{align}
\frac{ \kappa}{{c_{\rm s}} k_*}\cong \frac{ \sqrt{3} n w_* }{\varphi_*},
\end{align}
where $\varphi_*$ is the value of the scalar field at the crossing time, so $\beta$ is easily determined  by $\beta=w_*/\varphi_*^n$. The value of $w_*$ can be approximated by $w_0$ but it can be found more exactly   by solving  equation (\ref{mH}) at the horizon crossing time,  $\eta=\eta_*$.  We refer  the reader to \citep{Elizalde:2014uba}  for a comprehensive review.

\section{Parameter Estimation} \label{PS}
\noindent Now we investigate observational constraints on cosmological parameters of quasi-matter bounce cosmology. We use public package $\tt CosmoMC$ \citep{cosmomc} alongside the Planck measurement of CMB angular anisotropy in combination with the BICEP2/Keck Array (hereafter BK15) data \citep{BICEP2:2018kqh} to estimate Bayesian parameters. Here,  Planck data include both 2015 and 2018 measurements. 
Planck 2015 consists of $\rm {TT}$ angular power spectra for multipole $29 \leq l \leq 2508$ and  $\rm {lowP}$ which is the joint angular power spectra of $\rm {TT, EE, BB}$ and $\rm{TE}$ in multipole range $2 \leq l \leq 29$. Planck 2018 comes with $\rm {TT}$ and the joint $\rm{TT, TE, EE}$ angular power spectra for multipole $29 \leq l \leq 2508$, and $\rm {lowl}$, $\rm {lowE}$ for $\rm{TT}$ and $\rm{EE}$ angular power spectra for $2 \leq l \leq 29$. The main used cosmological parameters  are defined in Table \ref{mkh table 0}. 
\noindent Using Eqs. \ref{power spectrum}, \ref{Tensor_power spectrum}, we have  
\begin{equation} \label{mkh scalar spectra}
\mathcal{P}_{\rm s}(k) = A_{\rm s} \left(\frac{k}{k_*}\right)^{12 w_0} \exp{\left(\frac{\pi \hspace{1mm}\zeta}{2{c_{\rm s}} \hspace{1mm}k/k_*}\right)},
\end{equation}
and
\begin{equation}\label{mkh tensor spectra}
\mathcal{P}_{\rm t}(k) =A_{\rm t} \left(\frac{k}{k_*}\right)^{12 w_0} \exp{\left(\frac{\pi \hspace{1mm}\zeta}{2\hspace{1mm}k/k_*}\right)},
\end{equation}
\begin{table*}
	\begin{center}
		\small\addtolength{\tabcolsep}{-1pt}
		\begin{tabular}{ccccccccc}
			\noalign{\hrule\vskip 2pt} \noalign{\hrule\vskip 2pt}
			& 
			\multicolumn{5}{c}{Quasi-matter Bounce}&
			\multicolumn{1}{c}{$\Lambda$CDM} &
			\cr
			\omit&\multispan5\hspace{0.1cm}\hrulefill\hspace{0.1cm}&
			\multispan1\hspace{0.1cm}\hrulefill\hspace{0.1cm}
			\cr &
			\multicolumn{2}{c}{Planck 2015} & \multicolumn{3}{c}{Planck 2018} &
			\multicolumn{1}{c}{Planck 2018}
			\cr
			\omit&\multispan2\hspace{0.1cm}\hrulefill\hspace{0.1cm}&
			\multispan3\hspace{0.1cm}\hrulefill\hspace{0.1cm}&
			\multispan1\hspace{0.1cm}\hrulefill\hspace{0.1cm}
			\cr
			\noalign{\vskip 1pt}
			\multicolumn{1}{c}{Parameter} &  \multicolumn{1}{c}{$\rm TT$} & 
			\multicolumn{1}{c}{$\rm TT+BK15$}& \multicolumn{1}{c}{$\rm TT$}& 
			\multicolumn{1}{c}{$\rm TT + BK15$}& \multicolumn{1}{c}{$\rm TT, TE, EE + BK15$}&
			\multicolumn{1}{c}{$\rm TT, TE, EE$ \footnote{Here, TT, TE, EE denotes TT, TE, EE + lowE.}}
			\cr			
			\noalign{\hrule\vskip 1.5pt} \noalign{\hrule\vskip 1.5pt}
			$\Omega_{\mathrm{b}}h^2$ & $ 0.02237\pm 0.00025 $ & $0.02231\pm 0.00025$ & $0.02215\pm 0.00022$ & $0.02206\pm 0.00022$ & $0.02234\pm 0.00015$ & $0.02236 \pm 0.00015$
			\\
			$\Omega_{\mathrm{c}}h^2$& $ 0.1194\pm 0.0023 $ & $  0.1200\pm 0.0023$ & $0.1208\pm 0.0022 $ & $0.1219\pm 0.0021$ & $0.1208\pm 0.0014$ & $0.1202 \pm 0.0014$
			\\
			$100\theta_{\mathrm{MC}}$& $ 1.04096\pm 0.00049 $ & $1.04090\pm 0.00049$ & $1.04076\pm 0.00048$ & $1.04064\pm 0.00048$ & $1.04084\pm 0.00032$ & $1.04092 \pm 0.00031$
			\vspace*{.5mm} 
			\\
			$\ln(10^{10} \overline{A}_\mathrm{s})$ 
			& $ 3.138\pm 0.046 $ & $3.154\pm 0.046$ & 
			$3.052\pm 0.020$ &$3.057^{+0.017}_{-0.019}$ & $3.062^{+0.017}_{-0.020}$ & $3.045 \pm 0.016$
			\vspace*{.5mm}
			\\			
			$\tau$ & $ 0.0996\pm 0.024 $ & $ 0.106\pm 0.023$ & $0.0562\pm 0.0091$ & 
			$0.0570^{+0.0080}_{-0.0091}$ & $0.0604^{+0.0081}_{-0.0095}$ & $0.0544^{+0.0070}_{-0.0081}$
			\\
			$ {10}^{3} w_\mathrm{0}$\footnote{In quasi-matter bounce cosmology, $w_{\rm 0}$ plays the role of scalar spectral index $n_{\rm s}$ in $\Lambda$CDM through $n_{\rm s} -1 = 12 w_0 - \frac{\pi \zeta}{2 c_{\rm s}},$ see equation (\ref{spectral bounce}).}& $ -3.24\pm 0.58 $ & $-3.41\pm 0.57$ & $ -3.44\pm 0.56$ & $-3.72\pm 0.55$ & $-3.40\pm 0.44$ & $\dddot{}$
			\\
			$10^4\zeta$ & $ -3.40^{+3.15}_{-1.05}$ & $-2.45^{+1.80}_{-0.95}$ & $ -1.99^{+2.10}_{-0.68}$ & $-1.58^{+1.40}_{-0.77}$ & ${-1.67}^{+1.50}_{-0.83}$ & $\dddot{}$
			\vspace*{.5mm}
			\\
			$c_\mathrm{s}$ & $0.115\pm 0.048 $ & $0.087^{+0.035}_{-0.023}$ & $ 0.106\pm 0.045$ & $0.092^{+0.037}_{-0.024}$ & $0.097^{+0.037}_{-0.023}$ & $1$
			\vspace*{.5mm}
			\\
			\hline
			$H_0$ & $ 67.6\pm 1.0 \phantom{0}$ & 
			$67.3\pm 1.0  \phantom{0}$ & $66.85\pm 0.94 \phantom{0}$ & $ 66.34\pm 0.92 
			\phantom{0}$ & $67.04\pm 0.62$ & $67.27 \pm 0.60$
			\\
			$\Omega_{\mathrm{m}}$ & $ 0.312\pm 0.014 $ & ${0.316}^{+0.013}_{-0.015}$ & $0.322\pm 0.014$ & $0.3290\pm 0.0140$ & $0.3200\pm 0.0086$ & $0.3166 \pm 0.0084$
			\\
			$\sigma_8$& $ 0.846\pm 0.018 $ & $0.854\pm 0.017$&
			$ 0.8152\pm 0.0097 $ & $0.8200\pm 0.0094 $ & $0.8185\pm 0.0082$ & $0.8120 \pm 0.0073$ 
			\\
			\hline			
		\end{tabular}
	\end{center}
	\caption{
		Parameter 68\% intervals for the quasi-matter bounce and $\Lambda$CDM cosmology \citep{Akrami:2018odb} from Planck CMB angular power spectra with and without the $\rm{BK}15$ data. 
		\\			
		\textbf{Notes.} Here, for Planck 2015, $\rm TT$ stands for $\rm{TT + LowP}$, and for Planck 2018, $\rm TT$ is $\rm TT +\rm{lowl} + \rm {lowE}$, and $\rm{TT, TE, EE}$ 
		denotes  $ \rm {TT, TE, EE}  +\rm{lowl} + \rm {lowE}$.  
	}\label{mkh table 1}
\end{table*}
\begin{figure*}
	\centering
	\includegraphics[width=16cm]{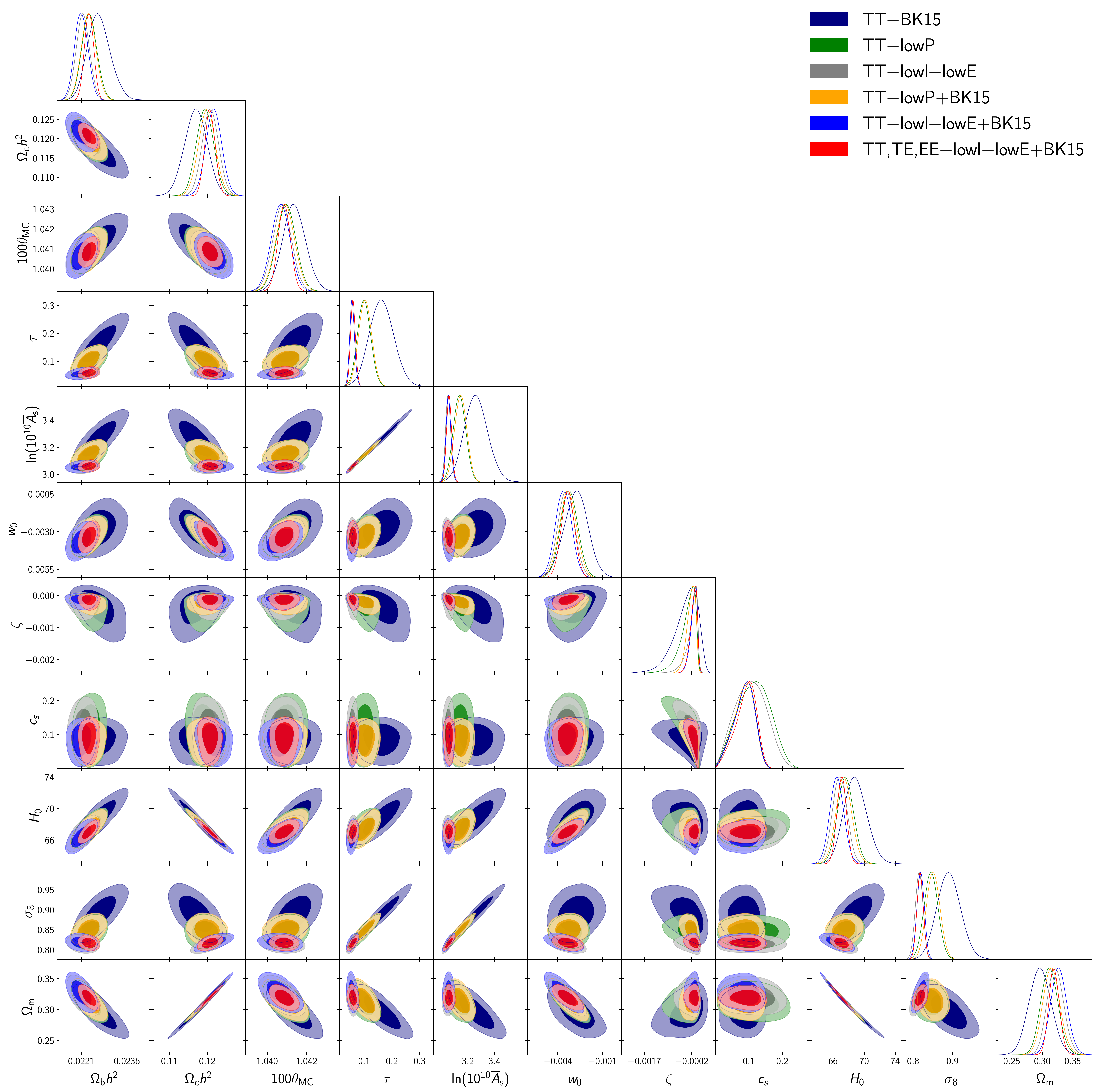}
	\caption
	{Marginalized joint 68\,\% and 95\,\%~CL regions for the cosmological parameters in quasi-matter bounce cosmology with  Planck TT+lowP (2015),
		and Planck TT (TT, TE, EE)+low+lowE (2018) in combination with the BK15 data.} \label{mkh fig 1}
\end{figure*}
for primordial scalar and tensor power spectrum, respectively, and the tensor-to-scalar ratio $r$
\begin{equation}
r \equiv \frac{\mathcal P_{\rm t}(k_*)}{\mathcal P_{\rm s}(k_*)} = 
\frac{\overline{A}_{\rm t}}{\overline{A}_{\rm s}}
= 24\hspace{.1mm}c^{3}_{\rm s}e^{\frac{\pi}{2} \zeta \left(1 - \frac{1}{c_{\rm s}}\right)} \cong 24\hspace{1mm}c^{3}_{\rm s},
\end{equation}
in which  $\overline {A}_{\rm s} \equiv \mathcal P_{\rm s} (k_*) = A_{\rm s} e^{\pi \zeta /2{c_{\rm s}}}$, $\overline {A}_{\rm t} \equiv \mathcal P_{\rm t} (k_*) = A_{\rm t} e^{\pi \zeta /2}$, and $\zeta = \xi/k_*$ is a dimensionless parameter. 
We have summarized parameters in Table \ref{mkh table 1}. {We break the degeneracy between $\zeta$ and $c_s$ in the argument of the exponential function of equation (\ref{mkh scalar spectra}) by using the tensor-to-scalar ratio, $r$. 
	This parameter examines the physical characteristics of primordial gravitational waves. The most prominent manifestation of these waves can be seen in CMB polarization $B$-modes caused by the scattering of an anisotropic CMB off of free electrons before decoupling occurs \citep{PhysRevD.55.1830}.
	Inclusion of BK15 data, which has a higher polarization sensitivity to the $B$-mode signal of primordial gravitational waves than Planck \citep{BICEP2:2018kqh,Akrami:2018odb}, strengthens the constraints on tensor-to-scalar ratio $24c^{3}_{s}$ and hence on $\zeta$.
	That is why $\zeta$ is getting smaller in magnitude when we include BK15 data.}

Because of the corresponding decrease in average optical depth $\tau$ in Planck 2018  with respect to Planck 2015, which is due to better noise sensitivity of the polarization likelihood ($\rm{lowE}$) than the joint temperature-polarization likelihood ($\rm {lowP}$) \citep{Ade:2015lrj,Akrami:2018odb}, power spectra parameters $(w_0,\zeta,{c_{\rm s}})$ are slightly smaller in Planck 2018 with respect to Planck 2015.  

We plotted the $68 \%$ and $95 \%$ confidence levels of the parameters in Fig. \ref{mkh fig 1}. {As we can see, aside from mathematical degeneracy between $\zeta$ and $c_s$, there is no correlation between $w_0$ and $c_{\rm s}$ and between $w_0$ and $\zeta$. Furthermore, except $w_0$ and $H_0$,  the free parameters of the bounce model do not have any significant correlation with other cosmological parameters. The correlation between $w_0$ and $H_0$ arises from the leading role of $w_0$ in the matter power spectrum and hence on matter density, which determines the expansion rate  $H(t)$.}

Investigating the physical effects of perturbation parameters ($w_0,\zeta,{c_{\rm s}}$) on the primordial scalar power spectrum requires a deep consideration of equation (\ref{mkh scalar spectra}). Current observational probes decisively rule out the scale-invariant power spectrum for primordial scalar fluctuations. In other words, the deviation from the scale-invariant power spectrum reveals the scalar spectral index and its scale dependency  $n_{\rm s}(k)$. 
Generally, the scale dependency of a power spectrum function can be represented by a Taylor expansion of the function with respect to $\ln k$. Therefore, the scalar power spectrum 
$\mathcal P_{s}(k) = A_s {\left(k/k_*\right)}^{n_{\rm s}-1}$ can be written as
\begin{align}
\ln {\mathcal P_{\rm s}(\ln k)}
&=  \ln A_{\rm s} + (n_{\rm s}-1)
[\ln (k/k_*)] \nonumber \\
&+\frac{1}{2}~\alpha_{\rm s}~
{[\ln (k/k_*)]}^2+ \frac{1}{6}~\beta_{\rm s}~
{[\ln (k/k_*)]}^3 + \cdot \cdot \cdot,
\end{align} 
in which second-order correction 
$\alpha_{\rm s} \equiv {d n_{\rm s}}/{d \ln k}$, 
and third-order correction  
$\beta_{\rm s} \equiv {d^2 n_{\rm s}}/{d \ln k^2}$ 
are running and running of the running of the scalar spectral index, respectively. This expression of the scalar power spectrum has some advantages.  
It shows that correction to the scale-invariant power spectrum is described in terms of the power of $\ln k$; the first correction indicates deviation from the scale invariance (primordial tilt), and the second correction takes into account scale dependency of the primordial tilt. Thus, for any arbitrary primordial scalar power spectrum  $\mathcal P_{s}( k)$, the first correction indicates the spectral index ($n_s -1$), and the second one denotes the running of the  spectral index $\alpha_s$
\begin{align}
n_{\rm s} -1 &\equiv  \frac{{\rm d} \ln\left[ \mathcal{P}_{\rm s}(\ln k)\right]}{{\rm d}\ln k}, \\
\alpha_{\rm s}  &\equiv 
\frac{{\rm d}^2 \ln\left[ \mathcal{P}_{\rm s}(\ln k)\right]}{{\rm d}\hspace*{1mm}{\ln k}^2}.
\end{align}
\noindent The new parameterized primordial scalar power spectrum obtained in section \ref{sec2} can be written as
\begin{align}
\ln \mathcal P_{\rm s} (\ln k) &= \ln A_{\rm s}+\frac{\pi \zeta}{2{c_{\rm s}}}+(12w_0-\frac{\pi \zeta}{2{c_{\rm s}}})\left[\ln(k/k_*)\right] \nonumber \\
&+\frac{1}{2}\frac{\pi \zeta}{2{c_{\rm s}}}{\left[\ln (k/k_*)\right]}^2 + \cdots ;
\end{align}
for $k=k_*$,  one can see that
$\,\ln \overline {A}_{\rm s}  = \ln A_{\rm s} +\pi \zeta /2{c_{\rm s}}$ as expected. Hence, the spectral index $n_{\rm Bs}$ and  running $\alpha_{\rm Bs}$ become 
\begin{align}
{n}_{\rm{Bs}} -1 &
= 12 \,w_0\,-\frac{\pi \, \zeta}{2 \,{c_{\rm s}}} \label{spectral bounce},\\ {\alpha}_{\rm {Bs}} &
=\frac{\pi \, \zeta}{2 \,{c_{\rm s}}}, \label{running bounce}
\end{align}
\begin{table*}
	\centering
	\begin{tabular}{cccc}
		\noalign{\hrule\vskip 2pt}
		\noalign{\hrule\vskip 3pt}
		Model & Parameter \hspace{4mm} & Planck TT+lowP \hspace{4mm} & Planck TT, TE, EE+lowl+lowE\\
		\hline
		& $n_{\mathrm{s}}$ & $0.9651 \pm 0.0066$ & $0.9635 \pm 0.0046$  \phantom{$\Big|$}\\
		$\Lambda$CDM+$\alpha_{\rm s}$ & $\alpha_\mathrm{s}$ & $-0.0084 \pm{0.0082}$ & $-0.0055 \pm{0.0067}$ \\
		\hline
		\multirow{1}{*}{}& $n_{\mathrm{Bs}}$ &
		$0.9636 \pm 0.0071$ & $0.9623 \pm 0.0055$ \phantom{$\Big|$}\\
		\multirow{1}{*}{$\rm{Quasi-matter\hspace{1mm} bounce\hspace{1mm}cosmology}$} 
		& $\alpha_\mathrm{Bs}$ &
		$-0.0036 \pm 0.0019$ & $-0.0021 \pm 0.0016$ \phantom{$\Big|$}\\
		\hline
	\end{tabular}
	\vspace{.4cm}
	\caption{\label{mkh table 2}
		Constraint on the primordial perturbation parameters for
		$\Lambda$CDM+$\alpha_{\rm s}$ \citep{Ade:2015lrj,Akrami:2018odb}, and quasi-matter bounce cosmology from {Planck} 2015 and 2018 CMB anisotropy measurements.}
\end{table*}\hspace{-2mm}where ``$\rm{Bs}$'' denotes the parameters in the  bouncing scenario. One may consider that there exists a conflict between  $\alpha_{\rm Bs}=(\pi/2) \; \zeta/c_{\rm s}$ and $\alpha_{\rm s}=2 \; \zeta/{c_{\rm s}}$ in equation (\ref{mkh alpha}). This discrepancy arises from ignoring terms of second- and higher-order approximations in the usual method and the linear approximation of the equation of state in our approach.    
It is worthwhile to mention that the tensor spectral indices,
\begin{equation}
{n}_{\rm{Bt}} -1 
= 12 \,w_0\,-\frac{\pi \zeta}{2}, \hspace{1cm} {\alpha}_{\rm {Bt}}
=\frac{\pi \zeta}{2},
\end{equation}
are different from the scalar ones because of the effect of the sound speed in the primordial scalar power spectrum.
\begin{figure}
	\includegraphics[width=8cm]{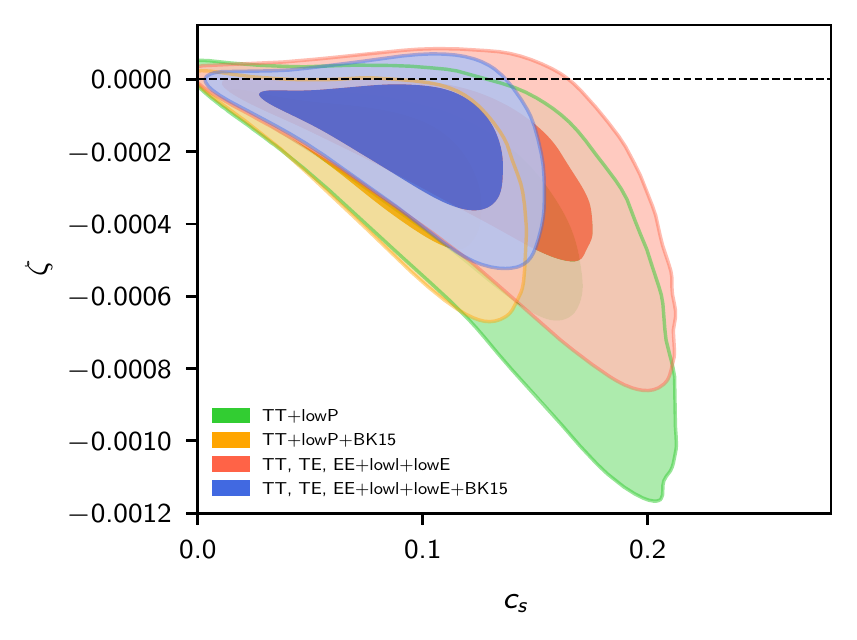}
	\includegraphics[width=8cm]{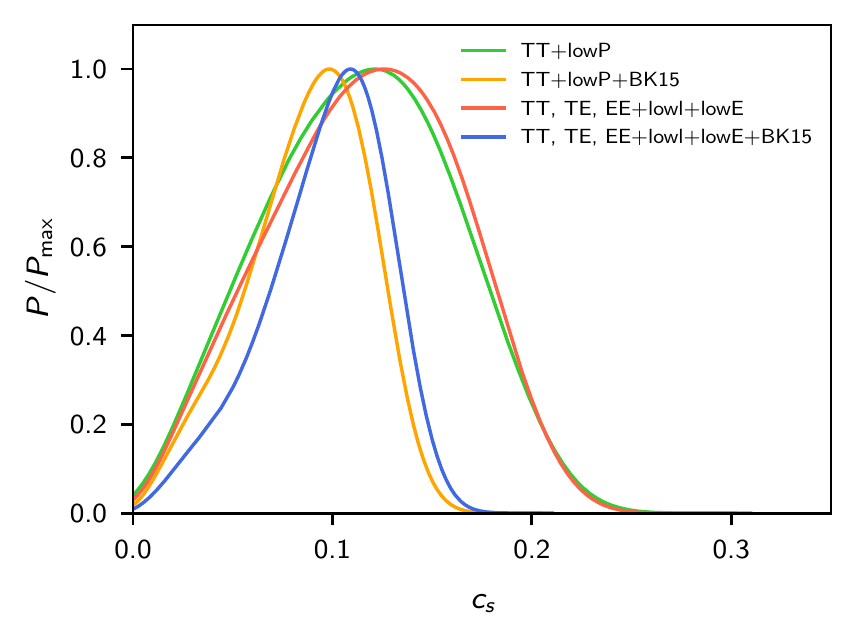}
	\caption
	{Marginalized joint 68\% and 95\% CL  for ($c_{\rm s}$,$\zeta$) and comparison of the posterior probability density of $c_{\rm s}$ in quasi-matter bounce cosmology, using Planck TT+lowP (2015) and Planck TT, TE, EE+lowl+lowE (2018) in combination with BK15 data. The dashed line indicates $\zeta=0$, which corresponds to the $\Lambda$CDM model.}
	\label{cs-xi fig 2}
\end{figure}
\begin{figure}
	\includegraphics[width=8cm]{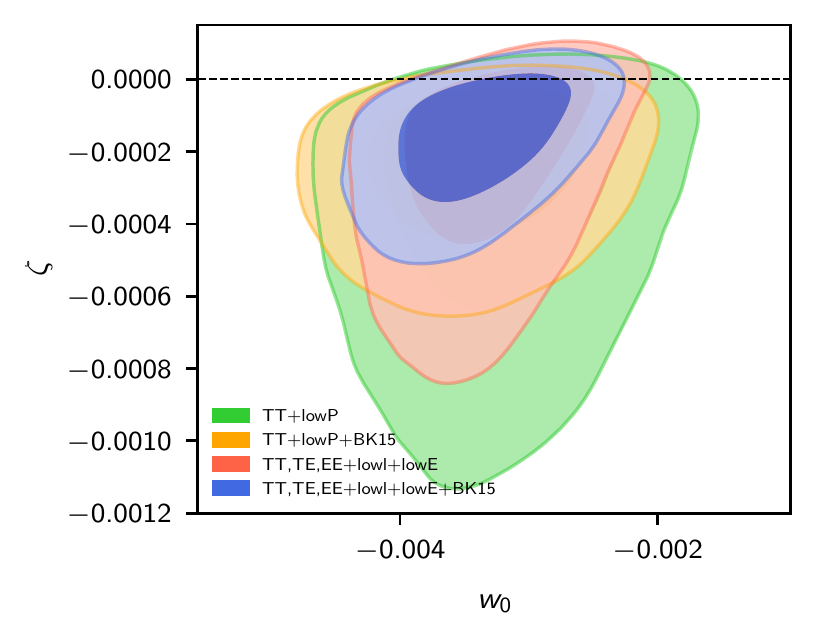}
	\includegraphics[width=8cm]{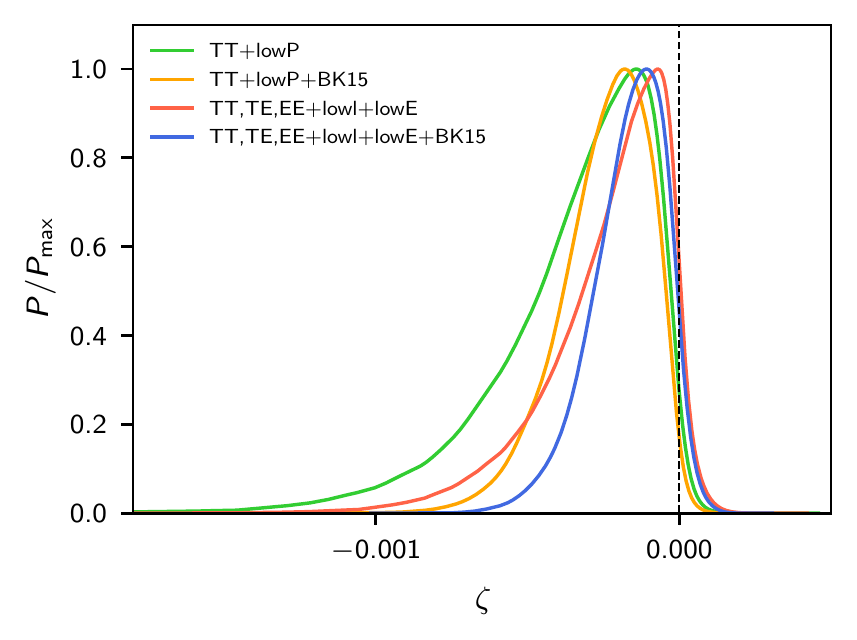}	
	\caption
	{Marginalized joint 68\% and 95\% CL  for ($w_0$,$\zeta$) and comparison of the posterior probability density of $\zeta$ in quasi-matter bounce cosmology, using Planck TT+lowP (2015) and Planck TT, TE, EE+lowl+lowE (2018) in combination with BK15 data. The dashed line indicates $\zeta=0$, which corresponds to $\Lambda$CDM model.} \label{mkh fig 2}
\end{figure}
Equation (\ref{running bounce}) confirms that a nonzero $\kappa$ gives rise to a scale dependent primordial tilt in quasi-matter bounce cosmology.{ The parameters have been listed in  Table \ref{mkh table 2}, and the joint constraints on ($c_{\rm s}$, $\zeta$) and ($w_0$, $\zeta$) are shown in Fig.\ref{cs-xi fig 2} and Fig.\ref{mkh fig 2}.} 
The Planck constraints on the $c_{\rm s}$ and $\zeta$ do not change significantly when complementary data set BK15 is included, but 
error bars are reduced. 
{For parameter $c_{\rm s}$, error bars are reduced by 30\% for Planck TT+lowl+lowE+BK15 with respect to Planck TT+lowl+lowE.}
We find at the $1\sigma$ confidence level 
\begin{align}
10^{4} \zeta = 
\left\{\begin{array}{lll}
-2.45^{+1.80}_{-0.95} && \text{(Planck TT+lowP+BK15)}
\\
\\
-1.67^{+1.50}_{-0.83} && \text{\small(Planck TT, TE,   EE+lowl+lowE+BK15)}
\end{array}\right.
\end{align}
which  means the slope $\kappa$ is non zero at the $1.4\sigma$ and 1.1$\sigma$ level for Planck TT+lowP+BK15 and  Planck TT+lowl+lowE+BK15, respectively, but gets close to zero at  2$\sigma$ confidence level for both Planck and Planck+BK15 data. 

\begin{figure}
	\includegraphics[width=8cm]{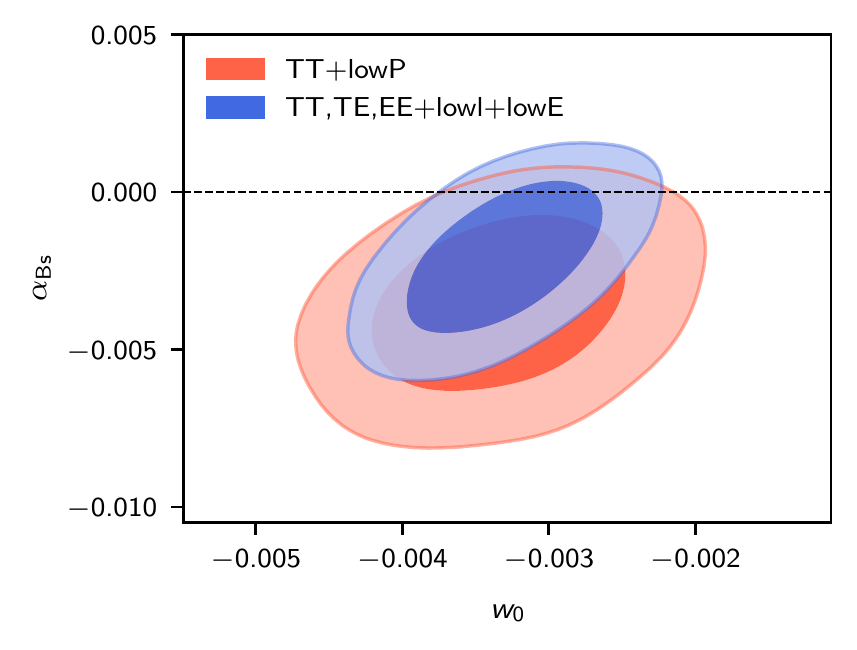}
	\includegraphics[width=8cm]{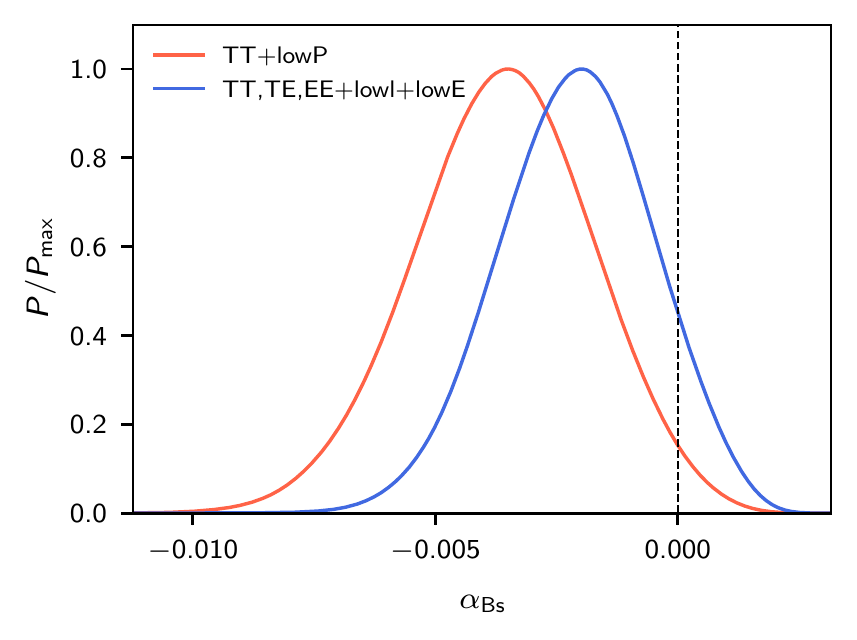}	
	\caption
	{Marginalized joint 68\% and 95\% CL for ($w_0$,$\alpha_{\rm Bs}$) and comparison of the posterior probability density of the running of the spectral index in quasi-matter bounce cosmology, using Planck TT+lowP (2015) and Planck TT, TE, EE+lowl+lowE (2018).} \label{mkh fig 3}
\end{figure}

\begin{figure}
	\includegraphics[width=8cm]{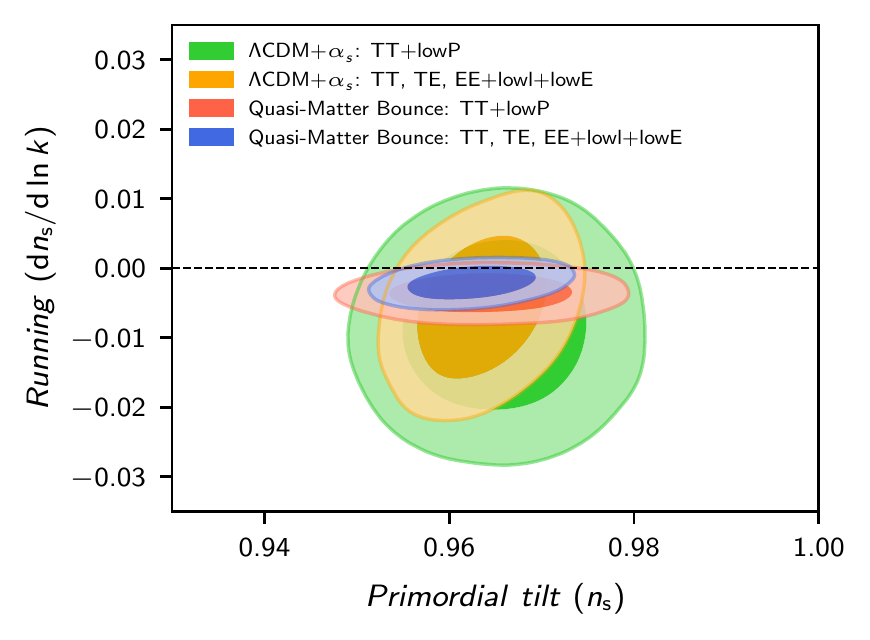}
	\includegraphics[width=8cm]{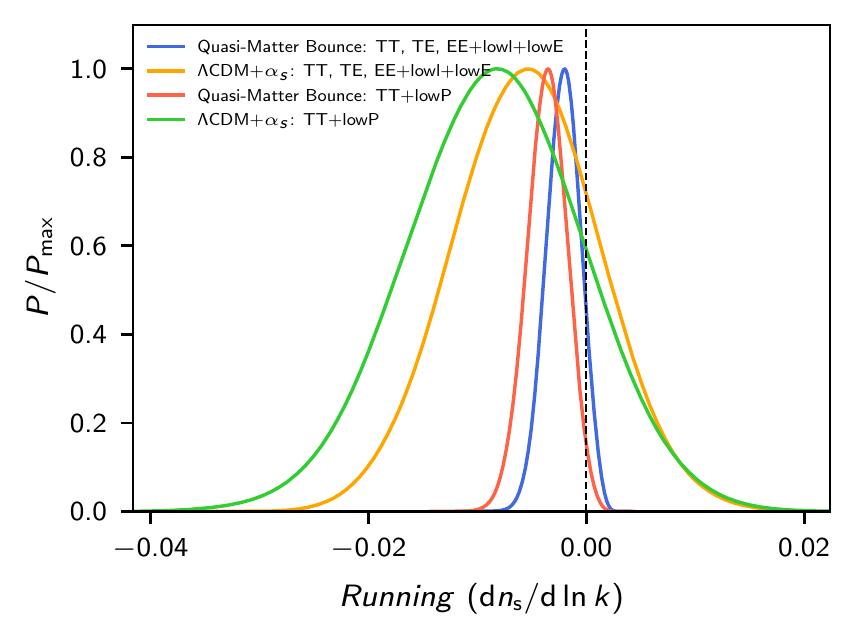}	
	\caption
	{Marginalized joint 68\% and 95\% CL for ($n_{\rm s}$,\,$ {\rm d} n_{\rm s}/ {\rm d} \ln k$) and comparison of the posterior probability density of the running of the scalar spectral index in $\Lambda$CDM+$\alpha_{\rm s}$ cosmology \citep{Ade:2015lrj,Akrami:2018odb} and the quasi-matter bounce scenario, using Planck TT+lowP (2015) and Planck TT, TE, EE+lowl+lowE (2018).} \label{mkh fig 4}
\end{figure}

{It is worth mentioning that, theoretically, the absolute value of the effective equation of state in the quasi-matter-dominated regime must be tiny, $|w|=|w_0+\kappa(\eta_* -\eta_0)|\ll1$. Furthermore,  $\kappa(\eta_* -\eta_0)$ as the first-order correction must be significantly smaller than $w_0$. It can be illuminated more precisely by using a rough approximation of the equation (\ref{mH}) at the crossing time
	\begin{equation}
	\kappa(\eta_* - \eta_0)\cong\frac{2\kappa}{\mathcal{H}_*} = \frac{\xi}{6 c_s k_*} = \frac{\zeta}{6 c_s} < w_0.
	\end{equation}
	Table \ref{mkh table 1} indicates that this condition is met, i.e. $(\zeta/6 c_s)/ w_0 \approx 10^{-1}$, which confirms that $w$ is a very small and slowly evolving value.}
The joint constraints on ($w_0$, $\alpha_{\rm Bs}$)  are  shown in Fig. \ref{mkh fig 3}. This implies that we obtain  at the $1\sigma$ confidence level 
\begin{align}
\alpha_{\rm {Bs}}  = 
\left\{ \begin{array}{rcl}
-0.0036 \pm 0.0019&&\text{(Planck TT+lowP)}
\\
\\
-0.0021 \pm 0.0016&&\text{\small (Planck TT, TE, EE+lowl+lowE)}
\\
\end{array}\right.
\end{align}
\hspace*{-2mm}which is nonzero at  the 2$\sigma$ and 1.3$\sigma$ level for Planck TT+lowP and Planck TT, TE, EE+lowl+lowE, respectively.  Error bars are reduced by $20\%$ compared to the {Planck TT+lowP}. Running  of the scalar spectral index for  Planck TT+lowP is still negative at the 2$\sigma$ confidence level but becomes positive for Planck TT, TE, EE+lowl+lowE. These results imply that  $\kappa$  leads to a scale dependent primordial tilt in  the primordial power spectrum for scalar fluctuations at the 1$\sigma$ confidence level. 
We do not claim that $\kappa$ gives rise to running of spectral index, because at the 2$\sigma$ confidence level there is no evidence for a scale dependent  tilt. Nevertheless, the  primordial scalar power spectrum in quasi-matter bounce cosmology provides a tighter constraint on the running of spectral index than the standard model of cosmology.  As shown in Fig. \ref{mkh fig 4}
the scalar spectral index in quasi-matter bounce cosmology is comparable to $\Lambda$CDM+$\alpha{\rm s}$, but running of the spectral index is more tightly constrained. Running  is nonzero at the $0.8 \sigma$ level in $\Lambda$CDM+$\alpha_{\rm s}$ for Planck TT, TE, EE+lowl+lowE, which is less constrained than quasi-matter bounce cosmology. 

\section{CONCLUSION}
\noindent We considered the influence of the linear time-dependent quasi-matter equation of state in the contracting phase of the universe in light of Planck CMB angular anisotropy measurements along with BICEP2/Keck Array data. We showed that the linear approximation of the equation of state gives rise to new primordial  spectra for scalar and tensor fluctuations. There is no need to define an extra parameter $\alpha_{\rm s}$  as  the running of the spectral index  for considering the scale dependency of the primordial tilt. In $\Lambda$CDM+$\alpha_{\rm s}$ cosmology, one needs a phenomenological generalization of the primordial scalar power spectrum to figure out the existence of a scale dependent scalar spectral index $n_{\rm s}$. In quasi-matter bounce cosmology, using this linear time-dependent equation of state without any need for such a generalization, one can investigate the existence of a running of the scalar spectral index. 

Significant constraints are the measurement of the linear equation-of-state parameters, applied in the new primordial power spectra, the zeroth-order approximation $w_0=-\,0.00340\pm 0.00044$, and the first-order correction $10^{4} \zeta= -1.67^{+1.50}_{-0.83}$ at the $1\,\sigma$ confidence level for  Planck TT, TE, EE+lowl+lowE+BK15. These amounts directly address quasi-matter field properties at the crossing time before the bounce. 
By expanding the new primordial scalar  power spectrum in terms of $\ln k$, we found out that the first-order correction $\kappa$ leads to running of the spectral index $\alpha_{\rm Bs}=\pi \zeta/2 {c_{\rm s}}$ at the $1\sigma$ confidence level. However, this result does not rule out zero running in quasi-matter models at the 2$\sigma$ or higher confidence level for Planck TT, TE, EE+lowl+lowE. Nevertheless, the running of the scalar spectral index in quasi-matter bounce cosmology is more tightly constrained than the standard model of cosmology. 

In quasi-matter bounce cosmology, running is determined to be $\alpha_{\rm Bs} = -\,0.0021 \pm 0.0016$ which is nonzero at the $1.3 \sigma$ level, whereas in $\Lambda$CDM+$\alpha_{s}$ it is nonzero at the $0.8 \sigma$ level for Planck TT, TE, EE+lowl+lowE. Similar to the $\Lambda$CDM, the scalar spectral index is determined to be  $n_{\rm Bs} = 0.9623 \pm 0.0055$, which lies $7 \sigma$ away from the scale-invariant primordial spectrum  for scalar perturbations for Planck TT, TE, EE+lowl+lowE.  

With the help of these new primordial spectra for the tensor and scalar perturbations, it was shown that the sound speed of primordial density fluctuations plays a crucial role in quasi-matter bounce cosmology. 
A nonunity value for the sound speed $c_{\rm s}$ implies that the spectral index and the running of the spectral index are different for the primordial scalar and tensor power spectra. Besides, Planck TT, TE, EE+lowl+lowE+BK15 yield that the sound speed of primordial density at the crossing time is  $c_{\rm s} = 0.097^{+0.037}_{-0.023}$ at the $1\sigma$ confidence level.

\section{ACKNOWLEDGMENTS}
We acknowledge support from the Institute for Advanced Studies in Basic Sciences (IASBS) for scientific computing time through the Gavazang Cluster. We are also grateful to Erfan Nourbakhsh and Ali Mansouri for stimulating discussions and valuable comments. Moreover, we thank Abdolhosein Khodam-Mohammadi, Ahmad Sheykhi, and Moein Mosleh for their support and suggestions.

\bibliography{ref}{}
\bibliographystyle{aasjournal}



\end{document}